\documentclass[aps,pra,nofootinbib,groupedaddress,twoside,twocolumn,final]{revtex4}
\usepackage{amsmath}
\usepackage{amsfonts}
\usepackage{amssymb}
\newtheorem{The}{Theorem}

\newtheorem{Lemma}[The]{Lemma}
\newtheorem{Pro}[The]{Proposition}

\newenvironment{proof}{\par\noindent\textit{Proof.\ }}{\hfill $\square$ \vspace{1em}}

\def\Cx{{\Bbb C}}\def\Rl{{\Bbb R}}\def\Nl{{\Bbb N}}\def\Ir{{\Bbb Z}}\def\idty{\openone}

\def\abs#1{\vert#1\vert}
\def\norm#1{\|#1\|}
\def\ketbra#1#2{\vert#1\rangle\langle#2\vert}
\def\ket#1{\vert#1\rangle}
\def\tr{{\rm tr}}
\newcommand{\kb}[1]{|#1\rangle\langle#1|}

\def\HH{{\cal H}}\def\B{{\cal B}}\def\A{{\cal A}}\def\M{{\cal M}}\def\C{{\cal C}}

\def\Hi{\HH_\infty} 
\def\Hii{\HH_{\infty\infty}} 
\newcommand{\scr}[1]{\mathcal{#1}}

\newcommand{\chan}{{\cal E}}

\def\weyl{{\mathbf W}}\def\Weyl{{\mathbf w}}
\def\alal{{\cal Z}}

\renewcommand\cosh{\rm ch}\renewcommand\sinh{\rm sh}

\begin{document}
\title{Infinitely entangled states}
\author{M. Keyl}
\email{M.Keyl@TU-BS.DE}
\affiliation{Inst. Math. Phys., TU-Braunschweig, Mendelssohnstra{\ss}e 3,
  D-38106 Braunschweig, Germany}
\author{D. Schlingemann}
\email{D.Schlingemann@TU-BS.DE}
\affiliation{Inst. Math. Phys., TU-Braunschweig, Mendelssohnstra{\ss}e 3,
  D-38106 Braunschweig, Germany}
\author{R.~F. Werner}
\email{R.Werner@TU-BS.DE}
\affiliation{Inst. Math. Phys., TU-Braunschweig, Mendelssohnstra{\ss}e 3,
  D-38106 Braunschweig, Germany}
\begin{abstract}
  For states in infinite dimensional Hilbert spaces
  entanglement quantities like the entanglement of distillation can become infinite.
  This leads naturally to the
  question, whether one system in such an infinitely
  entangled state can serve as a resource for tasks like the
  teleportation of arbitrarily many qubits. We show that appropriate states cannot be
  obtained by density operators in an infinite dimensional Hilbert space.
  However, using techniques for the description of infinitely many degrees of
  freedom from field theory and statistical mechanics, such states can nevertheless be constructed
  rigorously. We explore two related possibilities, namely an extended notion of
  algebras of observables, and the use of singular states on the algebra of bounded operators.
  As applications we construct the essentially unique infinite analogue of maximally entangled states,
  and the singular state used heuristically in the fundamental paper
  of Einstein, Rosen and Podolsky.
\\

\centerline{\it Dedicated to the memory of Rob Clifton}
\end{abstract}
\maketitle
\section{Introduction}

Many of the concepts of entanglement theory were originally developed for quantum systems described
in finite dimensional Hilbert spaces. This restriction is often justified, since we are usually
only trying to coherently manipulate a small part of the system. On the other hand, a full
description of almost any system, beginning with a single elementary particle, requires an infinite
dimensional Hilbert space. Hence if one wants to discuss decoherence mechanisms arising from the
coupling of the ``qubit part'' of the system with the remaining degrees of freedom, it is necessary
to widen the context of entanglement theory to infinite dimensions. This is not difficult, since
many of the basic notions, e.g. the definitions of entanglement measures, like the reduced von
Neumann entropy or entanglement of formation, carry over almost unchanged, merely with finite sums
replaced by infinite series. More serious are some technical problems arising from the fact that
such entanglement measures can now become infinite, and are no longer continuous functions of the
state. Luckily, as shown in recent work of Eisert et.~al.~\cite{EisSimPle01}, these problems can be
tamed to a high degree, if one imposes some natural energy constraints on the systems.

In the present paper we look at some not-so-tame states, which should be considered as idealized
descriptions of situations in which very much entanglement is available. For example, in the study
of ``entanglement assisted capacity'' \cite{BSST} one assumes that the communicating partners
have an unlimited supply of shared maximally entangled singlets. In quantum information problems
involving canonical variables it is easily seen that perfect operations can only be expected in the
limit of an ``infinitely squeezed'' two mode gaussian state as entanglement resource (see also
Section~\ref{sec:EPR}). But infinite entanglement is not only a
desirable resource, it is also a natural property of some
physical systems, such as the vacuum in quantum field theory (see \cite{SumWer87,SumWer88} and
Section~\ref{sec:maxim-entangl-stat} below). Our aim is to show that one can analyze these
situations by writing down bona fide states on suitably constructed systems.

It turns out that in order to do this we need to go one step beyond the standard Hilbert space
formalism of quantum mechanics (even with infinite dimensional Hilbert spaces). This is completely
analogous to other systems with infinitely many degrees of freedom, arising in quantum statistical
mechanics (in the thermodynamic limit) and quantum field theory. For example, consider a
translationally invariant, finite density equilibrium state of a gas. For such a state the
probability for having only finitely many particles must obviously be zero. In contrast, a state
given by density operators on Fock space implies a probability distribution of
particle numbers, which are by definition finite numbers in $\mathbb N$.
The expectation value of this distribution may be infinite. This means that in
the limit of taking many ($m\to\infty$) equally prepared systems, the total number $n$ of particles
diverges faster than $m$, so that $n/m\to\infty$. However, the particle number of each single
system remains finite.

Our first main result establishes a similar limitation for entanglement theory. Consider a state
with infinite distillible entanglement. This means that when we take a large number $m$ of copies
of such system pairs, and try to obtain from these a large number $n$ of nearly maximally entangled
singlet pairs by local operations and classical communication, we can achieve a rate $n/m\to\infty$
as $m\to\infty$. In this asymptotic sense each pair contains infinitely many ebits of entanglement.
But what does this mean for a single pair? Should we not get arbitrarily many singlets even out of
this? According to Theorem~\ref{nogo} the answer is no, as long as we stay in standard quantum
mechanics with bipartite states given by density operators on tensor product Hilbert spaces (of
finite or infinite dimension).

However, if we follow the lead of statistical mechanics, and employ the methods for describing
states with actually infinite particle number, we find a very natural framework to overcome this
limitation: Here we can have a direct mathematical representation of the intuitive idea of ``having
infinitely many singlets''. But we need to reconsider either the notion of states, allowing
``singular'' probability assignments on the space of quantum observables, which cannot be written
in terms of density operators (cf. Section~\ref{sec:singular-states}) or else allow more general
observable algebras for Alice and Bob (cf. Section~\ref{sec:von-neumann-algebras}). These two
approaches are closely related. We show in Section~\ref{sec:maxim-entangl-stat}, how this extended
framework leads to an essentially unique description of maximally entangled states of systems with
infinitely many degrees of freedom. In the final Section~\ref{sec:EPR} we discuss the ``original
EPR-state'', i.e., a mathematically rigorous version of the singular state employed by Einstein,
Rosen and Podolsky in their fundamental paper \cite{EPR35}. This Section builds on the work of Rob
Clifton \cite{CliftHalv98,CliftHalv99,ClHa01}, who sadly died while this paper was in preparation. We
dedicate it to his memory.

\section{Density operators on infinite dimensional Hilbert space}
\label{sec:dens-oper-infin}

We will start our discussion with a short look at entanglement properties of density operators on
an infinite dimensional but separable \footnote{Another
  extension of this framework, namely to
  Hilbert spaces of uncountable dimension (i.e., unseparable ones in
  the topological sense) is not really interesting with regard to
  entanglement theory, since any density operator has separable
  support, i.e., it is zero on all but countably many dimensions.}
Hilbert space $\HH \otimes \HH$.

Most of the definitions of entanglement quantities carry over from the finite dimensional setting
without essential change. Since we want to see how these quantities may diverge, let us look mainly
at the smallest, the distillible entanglement. It is defined as the largest rate $n/m$ of $n$
nearly perfect singlets, which can be extracted from $m$ pairs prepared in the given state by
protocols involving local quantum operations and classical communication. The class of quantum
operations is defined as in the finite dimensional case, by completely positive, trace preserving
maps (in the Schr{\"o}dinger picture). Effectively, distillation protocols for infinite dimensional
systems can be built up by first projecting to a suitable finite dimensional subspace, and
subsequently applying finite dimensional procedures to the result.

With this in mind we can easily construct {\it pure states} with infinite distillible entanglement.
Let us consider vectors in Schmidt form, i.e.,
 \begin{equation}\label{Schmidt}
  \Phi=\sum_n c_n\ e'_n\otimes e''_n\;
\end{equation}
with  orthonormal bases $e'_n$, $e''_n$, and positive numbers $c_n\geq0$, and $\sum_n\abs{c_n}^2=1$.
The density operator of the
restriction of this state to Alices's subsystem has eigenvalues $c_n^2$, and von Neumann entropy
$-\sum_n c_n^2\log_2(c_n^2)$, which we can take to be infinite ($c_n=1/(Z(n+2)\log_2(n+2)^2)$ will
do). We can distill this by using more and more of the dimensions as labelled by the bases
$e'_n,e''_n$, and applying the known finite dimensional distillation procedures to this to get out
arbitrary amount of entanglement per pair.

Once this is done, it is also easy to construct mixed states with large entanglement in the
neighborhood of any state $\rho$, mixed or pure, separable or not. We only have to remember that
every state is essentially (i.e., up to errors of small probability) supported on a finite
dimensional subspace. Therefore we can consider the mixture
$\rho_\epsilon=(1-\epsilon)\rho+\epsilon\sigma$ with a small fraction of an infinitely entangled
pure state $\sigma$, which is supported on those parts of Hilbert space, where $\rho$ is nearly
zero. Therefore distillation based on the support of $\sigma$ will work for $\rho_\epsilon$ and
produce arbitrarily large entanglement per $\rho_\epsilon$ pair, in spite of the constant reduction
factor $\epsilon$.

For the details of such arguments we refer to \cite{CliftHalv99,HorCirLew}. The argument as given
here does not quite show that states of infinite distillible entanglement are norm dense, but it
certainly establishes the discontinuity of the function ``distillible entanglement'' with respect
to the trace norm topology. This might appear to show that the approach to distillible entanglement
based on finite dimensional systems is fundamentally flawed: If only finitely many dimensions out
of the infinitely many providing a full description of the particle/system are used, might not the
entanglement be misrepresented completely? Here it helps that states living on a far out subspace
in Hilbert space usually also have large or infinite energy. For typical confined systems, the subspaces
with bounded energy are finite dimensional, so if we assume a realistic a priori bound on the
energy expectation of the states on the consideration, continuity can be restored
\cite{EisSimPle01}.

\section{Infinite one-copy entanglement}
\label{sec:infinite-one-copy}

If the entanglement of formation of a state is infinite: how much
of that entanglement can we get out? Since for pure states the
distillible entanglement is the same as the entanglement of
formation we know that given sufficiently many copies of the
state, we can use a distillation process producing in the long run
infinitely many nearly pure singlets per original entangled pair.
But if the entanglement is infinite, might it not be possible to
use only one copy of the state in the first place? In other words,
are there states, which can be used as a {\it one time resource},
to teleport an arbitrary number of qubits?

We will now give a definition of such states. The extraction of entanglement will be described by a
sequence of operations resulting in a pair of $d$-level systems with finite $d$. The extraction is
successful, if this pair is in a nearly maximally entangled state, when one starts from the given
input state. The overall operation is then given mathematically by a completely positive, trace
preserving map $\chan_d$. Of course, we must make sure that the extraction process does not
generate entanglement. There are different ways of expressing this mathematically. For example, we
could allow $\chan_d$ to be composed of an arbitrary number of rounds of local quantum operations
and classical communication (``LOCC operations''). We will also consider a much weaker, and much
more easily verified condition, namely that $\chan_d$ takes pure
product states into states with positive partial transpose (``PPtPPT operations'' for ``pure
product to positive partial transpose''). Of course, every LOCC channel is a PPtPPT channel.

The success is measured by the fidelity (overlap) of the output state $\chan_d(\rho)$ with a fixed
maximally entangled state on $\Cx^d\otimes\Cx^d$. By $p_d$ we denote the projection onto this
maximally entangled reference vector. Then a density operator $\rho$ is said to have  ``infinite
one-copy entanglement", if for any \mbox{$\varepsilon>0$} and any $d\in\Bbb{N}$ there is a PPtPPT
channel $\chan_d$  such that
\begin{equation}\label{infentR}
\tr(\chan_d(\rho) p_d) \geq1-\varepsilon\;.
\end{equation}
Then we have the following Theorem, whose proof uses a distillation estimate of Rains
\cite{Rain00} developed for the finite dimensional context.

\begin{The}\label{nogo}
For any sequence of PPtPPT channels $\chan_d$, $d \in \Bbb{N}$, and for any fixed density operator
$\rho$ we have
\begin{equation}\label{limNogo}
  \lim_{d\to\infty}\tr(\chan_d(\rho) p_d)=0\;.
\end{equation}
In particular, no density operator with infinite one-copy entanglement exists.
\end{The}
\begin{proof}
Consider the operators $A_d$ defined by
\begin{equation}\label{channel-pptppt}
  \tr(\rho A_d)=\tr(\chan_d(\rho)p_d).
\end{equation}
In order to verify that $A_d$ exists, observe that $\chan_d$,
as a positive operator is automatically norm continuous. Hence the
right hand side is a norm continuous linear functional on density matrices $\rho$. Since the set of
bounded operators is the dual Banach space of the set of trace class operators \cite[Theorem VI.26]{RESI1}
such functionals are indeed of the form (\ref{channel-pptppt}).
We now have to show that, for every $\rho$, we have
$\lim_d\tr(\rho A_d)=0$, i.e., that $A_d\to0$  in the weak*-topology of this dual Banach space.

Obviously, $0\leq A_d\leq\idty$, and by the Banach-Alaoglu Theorem \cite[Theorem IV.21]{RESI1}, this set is
compact in the topology for which we want to show convergence. Hence the sequence has accumulation
points and we only have to show that all accumulation points are zero. Let $A_\infty$ denote such a
point. Then it suffices to show that $\tr(\sigma A_\infty)=0$ for all pure product states $\sigma$.
Indeed, since $A_\infty\geq0$, this condition forces $A_\infty \phi\otimes\psi=0$ for all pairs of
vectors $\phi,\psi$, and hence  $A_\infty=0$, because such vectors span the tensor product Hilbert
space.

On the other hand, our locality condition is strong enough to allow us to compute the limit
directly for pure product states $\sigma$. We claim that
\begin{eqnarray}
 \tr(\sigma A_d)
   &=&\tr(\chan_d(\sigma)p_d) \nonumber\\
   &=&\tr(\chan_d(\sigma)^{T_2}p_d^{T_2})
      \leq \|p_d^{T_2}\|
   =1/d
\end{eqnarray}
Here we denote by $X^{T_2}$ the partial transposition with respect to the second tensor factor, of
an operator $X$ on the finite dimensional space $\Cx^d\otimes\Cx^d$, and use that this operation is
unitary with respect to the Hilbert-Schmidt scalar product $\langle X,Y\rangle_{HS}=\tr(X^*Y)$. By
assumption, $\chan_d(\sigma)^{T_2}\geq0$, and since partial transposition preserves the trace,
$\chan_d(\sigma)^{T_2}$ is even a density operator. Hence the expectation value of $p_d^{T_2}$ in
this state is bounded by the norm of this operator. But it is easily verified that $p_d^{T_2}$ is
just $(1/d)$ times the unitary operator exchanging the two tensor factors. Hence its norm is
$(1/d)$. Taking the limit of this estimate along a sub-net of $A_d$ converging to $A_\infty$, we
find $\tr(\sigma A_\infty)=0$.
\end{proof}

\section{Singular states and infinitely many degrees of freedom}
\label{sec:sing-stat-infin}

In this section we will show how to construct a system of infinitely many singlets.  It is clear from
Theorem~\ref{nogo} that not all of the well-known features of the finite situation will
carry over. Nevertheless, we will stay as closely as possible to the standard constructions trying
to pretend that $\infty$ is finite, and work out the necessary modifications as we go along.

\subsection{Von Neumann's incomplete infinite tensor product of Hilbert spaces}

The first difficulty we encounter is the construction of Hilbert spaces for Alice's and Bob's
subsystem, respectively, which should be the {\it infinite tensor power} $(\Cx^2)^{\otimes \infty}$
of the one qubit space $\Cx^2$. Let us recall the definition of a tensor product: it is a Hilbert
space generated by linear combination and norm limits from basic vectors written as
$\Phi=\bigotimes_{j=1}^\infty\phi_j$, where $\phi_j$ is a vector in the $j$th tensor factor. All we
need to know to construct the tensor product as the completion of formal linear combinations of
such vectors are their scalar products, which are, by definition,
\begin{equation}\label{infsp}
  \left\langle\bigotimes_{j=1}^\infty\phi_j,\
      \bigotimes_{j=1}^\infty\psi_j\right\rangle
  = \prod_{j=1}^\infty\langle\phi_j,\psi_j\rangle\;.
\end{equation}
The problem lies in this infinite product, which clearly need not converge for arbitrary choice of
vectors $\phi_j,\psi_j$. A well-known way out of this dilemma, known as {\it von Neumann's
incomplete tensor product} \cite{vNeuP} is to restrict the possible sequences of vectors
$\phi_1,\phi_2,\ldots$ in the basic product vectors: for each tensor factor, one picks a reference
unit vector $\chi_j$, and only sequences are allowed for which $\phi_j=\chi_j$ holds for all but a
finite number of indices. Evidently, if this property holds for both the $\phi_j$ and the $\psi_j$
the product in (\ref{infsp}) contains only a finite number of factors $\neq1$, and converges. By
taking norm limits of such vectors we see that also product vectors for which
$\sum_{j=1}^\infty\norm{\phi_j-\chi_j}<\infty$ are included in the infinite product Hilbert space.
However, the choice of reference vectors $\chi_j$ necessarily breaks the full unitary symmetry of
the factors, as far as asymptotic properties for $j\to\infty$ are concerned. For the case at hand,
i.e., qubit systems, let us choose, for definiteness, the ``spin up'' vector as $\chi_j$ for every
$j$, and denote the resulting space by $\Hi$.

An important observation about this construction is that all observables of finite tensor product
subsystems act as operators on this infinite tensor product space.
In fact, any operator $\bigotimes_{j=1}^\infty A_j$ makes
sense on the incomplete tensor product, as long as $A_j=\idty$ for all but finitely many indices.
The algebra of such operators is known as the algebra of local observables. It has the structure of
a *-algebra, and its closure in operator norm is called  {\it quasi-local algebra} \cite{BraRob}.

Let us take the space $\Hi$ as Alice's and Bob's Hilbert space.
Then each of them holds infinitely many qubits, and we can discuss
the entanglement contained in a density operator on
$\Hi\otimes\Hi$. Clearly, there is no general upper bound to this
entanglement, since we can take a maximally entangled state on the
first $M<\infty$ factors, complemented by infinitely many spin-up
product states on the remaining qubit pairs. But for any fixed
density operator the entanglement is limited: for measurements on
qubit pairs with sufficiently large $j$ we always get nearly the
same expectations as for two uncorrelated spin-up qubits (or
whatever the reference states $\chi_j$ dictate). This is just
another instance of Theorem~\ref{nogo}: there is no density
operator describing infinitely many singlets.

\subsection{Singular states}
\label{sec:singular-states}

However, can we not take the limit of states with growing entanglement? To be specific, let
$\Phi_M$ denote the vector which is a product of singlet states for the first $M$ qubit pairs, and
a spin-up product for the remaining ones. These vectors do not converge in $\Hi\otimes\Hi$, but
that need not concern us, if we are only interested in expectation values: for all local
observables $A$ (observables depending on only finitely many qubits), the limit
 \begin{equation}\label{omegai}
  \omega(A)=\lim_M\langle{\Phi_M,A\Phi_M}\rangle
\end{equation}
exists. Thereby we get an expectation value functional for all quasi-local observables, and by the
Hahn-Banach Theorem (see e.g. \cite[Theorem III.6]{RESI1}),
we can extend this expectation value functional to all bounded operators on
$\Hi\otimes\Hi$. The extended functional $\omega$ has all the properties required by the
statistical interpretation of quantum mechanics: linearity in $A$, $\omega(A)\geq0$ for positive
$A$, and $\omega(\idty)=1$. In the terminology of the theory of operator algebras, it is a {\em state}
on the algebra of all bounded operators. By construction, $\omega$ describes maximal entanglement for
{\it any\/} finite collection of qubit pairs, so it is truly a state of infinitely many singlets.

How does this match with Theorem~\ref{nogo}? The crucial point is that that Theorem only speaks of
states given by the trace with a density operator, i.e., of functionals of the form
$\omega_\rho(A)=\tr(\rho A)$. Such states are called ``normal''. But there is no density operator
for $\omega$: this is a {\it singular state} on the algebra of bounded operators.

Singular states are not that unusual in quantum mechanics, although they can only be
``constructed'' by an invocation the Axiom of Choice, usually through the Hahn-Banach Theorem
{ \footnote{Other constructions based on the Axiom of Choice are the application of invariant means,
 e.g., when averaging expectation values over all translations, or algebraic constructions using
 maximal ideals. For an application in von Neumann style measurement theory of continuous spectra,
 see \cite{Oza}}.
For example, we can think of a non-relativistic particle localized at a sharp point, as witnessed
by the expectations of all continuous functions of position. Extending from this algebra to all
bounded operators, we get a singular state with sharp position
 \footnote{This is not related to improper eigenkets of position, which do not yield normalized
 states},
but ``infinite momentum'', i.e., the probability assigned to finding the momentum in any given
finite interval is zero \cite{QHA2}. This shows that the probability measure on the momentum space
induced by such a state is only finitely additive, but not $\sigma$-additive. This is typical for
singular states.

More practical situations involving singular states arise in all systems with infinitely many
degrees of freedom, as in quantum field theory and in statistical mechanics in the thermodynamic
limit. For example, the equilibrium state of a free Bose gas in infinite space at finite density
and temperature is singular with respect to Fock space because the probability for finding
only a finite number of particles in such a state is zero. In all these cases, one is primarily
interested in the expectations of certain meaningful observables (e.g., local observables), and the
wilder aspects of singular states are connected only to the extension of the state to {\it all}
bounded operators. Therefore it is a good strategy to focus on the state as an expectation
functional only on the ``good'' observables.

\subsection{Local observable algebras}
\label{sec:von-neumann-algebras}

If we want to represent a situation with infinitely many singlets, an obvious approach is to take
again  von Neumann's incomplete tensor product, but this time the infinite tensor product of {\it
pairs} rather than single qubits, with the singlet vector chosen as the reference vector $\chi_j$
for every pair. We denote this space by $\Hii$, and by  $\Omega\in\Hii$  the infinite tensor
product of singlet vectors. Clearly, this is a normal state (with density operator
$\ketbra\Omega\Omega$), and we seem to have gotten around Theorem~\ref{nogo} after all.

However, the problem is now to identify the Hilbert spaces of Alice and Bob as tensor factors of
$\Hii$. To be sure, the observables measurable by Alice and Bob, respectively, are easily
identified. For example, the $\sigma_x$-Pauli matrix for Alice's 137$^{\rm th}$ particle is a well defined
operator on $\Hii$. Alice's observable algebra $\A$ is generated by the collection of all Alice
observables for each pair. Bob's observable algebra $\B$ is generated similarly, and together they
generate the local algebra of the pair system. Moreover, the two observable algebras commute
elementwise. This is just what we expect from the usual setup, when the total Hilbert space is
$\HH=\HH_A\otimes\HH_B$, and Alice's and Bob's observable algebras are $\A=\B(\HH_A)\otimes\idty_B$
and $\B=\idty_A\otimes\B(\HH_B)$.

However, the $\A$ and $\B$ constructed above are definitely not of this form, so $\Hii$ has no
corresponding decomposition as $\HH_A\otimes\HH_B$. The most direct way of seeing this is to note
that $\Hii$ contains no product vectors describing an uncorrelated preparation of the two
subsystems. If we move to qubit pairs with sufficiently high index, then by construction of the
incomplete tensor product, {\it every} vector in $\Hii$ will be close to the singlet vector, and in
particular, will violate Bell's inequality nearly maximally (see also Section~\ref{sec:Bellmax}).

Hence we arrive at the following generalized notion of bipartite states, generalizing the finite
dimensional one: Alice's and Bob's subsystems are identified by their respective observable
algebras $\A$ and $\B$. We postpone the discussion of the precise technical properties of these
algebras. What is important is, on the one hand, that these algebras are part of a larger system,
so they are both subalgebras of a larger algebra, typically the algebra $\B(\HH)$ of bounded
operators on some Hilbert space. This allows us to consider products and correlations between the
two algebras. On the other hand, each measurement Alice chooses must be compatible with each one
chosen by Bob. This requires that $\A$ and $\B$ commute elementwise. A {\it bipartite state} is
then simply a state on the algebra containing both $\A$ and $\B$.

We can then describe the two ways out of the NoGo-Theorem: on the one hand we can allow more
general states than density matrices, but on the other hand we can also consider more general
observable algebras. In the examples we will discuss, the algebra containing $\A$ and $\B$ will in
fact be of the form $\B(\HH)$, and the states will be given by density matrices on $\HH$. So both
strategies can be successful by themselves.

\subsection{Some basic facts about operator algebras}

The possibility of going either to singular states or to extended observable algebras is typical of
the duality of states and observables in quantum mechanics. There are many contexts, where it is
useful to extend either the set of states or the set of observables by idealized elements, usually
obtained by some limit. However, these two idealizations may not be compatible \cite{QHA2}. There
are two types of operator algebras which differ precisely in the strength of the limit procedures
under which they are closed \cite{BraRob,Tak}.

On the one hand there are \emph{C*-algebras}, which are isomorphic to norm and adjoint closed
algebras of operators on a Hilbert space. Norm limits are quite restrictive, so some operations are
not possible in this framework. In particular, the spectral projections of an hermitian element of
the algebra often do not lie again in the algebra (although all continuous functions will).
Therefore, it is often useful to extend the algebra by all elements obtained as weak limits
(meaning that all matrix elements converge). In such \emph{von Neumann algebras} the spectral
theorem holds. Moreover, the limit of an increasing but bounded sequence of elements always
converges in the algebra. For these algebras the distinction between normal and singular states
becomes relevant. The normal states are simply those for which such increasing limits converge, and
at the same time those which can be represented by a density operator in the ambient Hilbert space.

A basic operation for von Neumann algebras is the formation of the {\it commutant}: for any set
$\M\subset\B(\HH)$ closed under the adjoint operation, we define its commutant as the von Neumann
algebra
\begin{equation}\label{commutant}
  \M'=\Bigl\{ X\in\B(\HH)\Bigm\vert
             \forall M\in\M\ [M,X]=0\Bigr\}\;.
\end{equation}
Then the Bicommutant Theorem \cite{Tak} states that $\M''=(\M')'$ is the smallest von Neumann
algebra containing $\M$. In particular, when $\M$ is already an algebra, $\M''$ is the weak closure
of $\M$.  Von Neumann algebras are characterized by the property $\M''=\M$. A von Neumann algebra
$\M$ with the property that its only elements commuting with all others are the multiples of the
identity (i.e., $\M'\cap\M''=\Cx\idty$) is called a {\it factor}.

It might seem that the two ways out of the NoGo-Theorem indicated at the end of the previous
section are opposite to each other, but in fact they are closely related. For if $\omega$ is a
state on a C*-algebra $\C\supset\A\cup\B$, we can associate with it a Hilbert space $\HH_\omega$, a
representation $\pi_\omega:\C\to\B(\HH)$, and a unit vector $\Omega\in\HH_\omega$, such that
\mbox{$\omega(C)=\langle\Omega,\pi_\omega(C)\Omega\rangle$}, and such that the vectors
$\pi_\omega(C)\Omega$ are dense in $\HH_\omega$. This is called the Gelfand-Naimark-Segal
(GNS)-construction \cite{BraRob}. Clearly, the given state $\omega$ is given by a density operator
(namely $\ketbra\Omega\Omega$) in this new representation and the algebra can naturally be extended
to the weak closure $\pi_\omega(C)''$. The commutativity of two subalgebras is preserved by the
weak closure, so the normal state $\ketbra\Omega\Omega$, and the two commuting von Neumann
subalgebas $\pi_\omega(\A)''$ and $\pi_\omega(\B)''$ are again a bipartite system, which describes
essentially the same situation. The only difference is that some additional idealized observables
arise from the weak closure operations, and that some observables in $\C$ (those with $C\geq0$ but
$\omega(C)=0$) are represented by zero in $\pi_\omega$.

We remark that von Neumann's incomplete infinite tensor product of Hilbert spaces can be seen as a
special case of the GNS-construction: The infinite tensor product of C*-algebras $\bigotimes_i\A_i$
is well-defined (see \cite[Sec 2.6]{BraRob} for precise conditions), essentially by taking the norm
completion of the algebra of \emph{local observables} $\bigotimes_iA_i$, with all but finitely many
factors $A_i\in\A_i$ equal to $\idty_i$. On this algebra the infinite tensor product of states is
well-defined, and we get the incomplete tensor product as the GNS-Hilbert space of the algebra
$\bigotimes_i\B(\HH_i)$ with respect to a the pure product state defined by the reference vectors
$\chi_i$.

\section{Von Neumann algebras with Maximal entanglement}
\label{sec:maxim-entangl-stat}
\subsection{Characterization and basic properties}
\label{sec:maxent}

Let us analyze the example given in the last section: the bipartite state obtained from the
incomplete tensor product of singlets in $\Hii$. We take as Alice's observable algebra $\A$ the von
Neumann algebra generated by all local Alice operators (and analogously for $\B$ob). The bipartite
state on these algebras, given by the reference vector $\bigotimes_i\chi_i$, then has the following
properties

\begin{description}
 \item[\rm ME 1] \label{item:1} $\A$ and $\B$ together \emph{generate} $\B(\HH)$ as a von Neumann algebra,
   so there are no additional observables of the system beyond those measurable by Alice and Bob.
 \item[\rm ME 2] \label{item:2} $\A$ and $\B$ are \emph{maximal} with respect to mutual commutativity.
   (i.e., $\A=\B'$ and $\B=\A'$)
 \item[\rm ME 3] The overall state is \emph{pure}, i.e., given by a vector $\Omega\in\HH$,
 \item[\rm ME 4] \label{item:3} The restriction of this state to either subsystem is a \emph{trace}, so
   $\omega(A_1A_2)=\omega(A_2A_1)$, for $A_1,A_2\in\A$.
 \item[\rm ME 5] $\A$ is \emph{hyperfinite}, i.e., it is the weak closure of an increasing family of
    finite dimensional algebras.
\end{description}

These properties, except perhaps ME 2 (see \cite{AraWoo}) are immediately clear from the
construction, and the properties of the respective local observables. They are also true for finite
dimensional maximally entangled states on $\HH=\HH_A\otimes\HH_B$, $\A=\B(\HH_A)\otimes\idty$, and
$\B=\idty\otimes\B(\HH_B)$. This justifies calling this particular bipartite system \emph{maximally
entangled}, as well.

There are many free parameters in this construction. For example, we could take arbitrary
dimensions $d_i<\infty$ for the $i^{\rm th}$ pair. However, all these possibilities lead to the
same maximally entangled system:

\begin{The}\label{maxentio}
All bipartite states on infinite dimensional systems satisfying conditions {\rm ME 1 - ME 5}
above are unitarily
isomorphic.
\end{The}

\begin{proof} (Sketch). We first remark that $\A$ has to be a {\it factor}, i.e.,
$\A\cap\A'=\Cx\idty$. Indeed, using ME 1 and ME 2, we get
$\A\cap\A'=\B'\cap\A'=(\B\cup\A)'=\B(\HH)'=\Cx\idty$.

Now consider the support projection $S\in\A$ of the restriction of the state to $\A$. Thus
$\idty-S$ is the largest projection in $\A$ with vanishing expectation. Suppose that this
projection does not lie in the center of $\A$, i.e., there is an $A\in\A$ such that $AS\neq SA$.
Let $X=(\idty-S)AS$, which must then be nonzero, as $AS-SA=((\idty-S)+S)(AS-SA)=X-SA(\idty-S)$.
Then using the trace property we get $\omega(X^*X)=\omega(XX^*)\leq\norm A^2\omega(\idty-S)=0$,
which implies that the support projection of $X^*X$ has vanishing expectation. But since
$X^*X\leq\norm A^2S$, this contradicts the maximality of $(\idty-S)$. It follows that $S$ lies in
the center of $\A$ and that $S=\idty$, because $\A$ is a factor. To summarize this argument,
$\omega$ must be {\it faithful}, in the sense that $A\in\A$, $A\geq0$, and $\omega(A)=0$ imply
$A=0$.

Now consider the subspace spanned by all vectors of the form $A\Omega$, with $A\in\A$. This
subspace is invariant under $\A$, so its orthogonal projection is in $\A'=\B$. But since
$(\idty-P)$ obviously has vanishing expectation, the previous arguments, applied to $\B$ imply that
$P=\idty$. This is to say that $\A\Omega$ is dense in $\HH$ or, in the jargon of operator algebras,
that $\Omega$ is {\it cyclic} for $\A$. Thus $\HH$ is unitarily equivalent to the GNS-Hilbert space
of $\omega$ restricted to $\A$, and the form of $\B=\A'$ is completely determined by this
statement. Now a factor admits at most one trace state, so $\omega$ is uniquely determined by the
isomorphism type of $\A$ as a von Neumann algebra, and it remains to show that $\A$ is uniquely
determined by the above conditions.

$\A$ is a factor admitting a faithful normal trace state, so it is a ``type II$_1$-factor'' in von
Neumann's classification. It is also hyperfinite, so we can invoke a deep result of Alain Connes
\cite{connes} stating that such a factor is uniquely determined up to isomorphism.\end{proof}

For the rest of this section we will study further properties of this unique maximally entangled
state of infinite entanglement. The items ME 6, ME 7 below are clear from the above proof. ME 8 follows by
splitting the infinite tensor product either into a finite product and an infinite tail, or into
factors with even and odd labels, respectively.  ME 9 - ME 11 are treated in separate subsections as
indicated.

\begin{description}
 \item[\rm ME 6] $\A$ and $\B$ are factors: $\A\cap\A'=\Cx\idty$.
 \item[\rm ME 7] $\A\Omega$ and $\B\Omega$ are dense in $\HH$.
 \item[\rm ME 8] The state contains infinite one-shot entanglement, which is not
 diminished by extracting entanglement.
   Moreover, it is unitarily isomorphic to two copies of itself.
 \item[\rm ME 9] Every density operator on $\HH$ maximally violates the Bell-CHSH inequality
  (see Section~\ref{sec:Bellmax}).
 \item[\rm ME 10] The generalized Schmidt spectrum of $\Omega$ is flat
 (see Section~\ref{sec:schm-decomp-modul}).
  \item[\rm ME 11] Every $A\in\A$ is completely correlated with a ``double'' $B\in\B$.
  (see Section~\ref{sec:EPR-prop}).
\end{description}

\subsection{Characterization by violations of Bell's inequalities}
\label{sec:Bellmax}

If we look at systems consisting of two qubits, maximally entangled states can be characterized in
terms of maximal violations of Bell-inequalities. It is natural to ask, whether something similar
holds for the infinite dimensional setting introduced in Section \ref{sec:maxim-entangl-stat}. To
answer this question consider again a bipartite state $\omega$ on an algebra containing  two
mutually commuting  algebras $\scr{A}, \scr{B}$ describing Alice's and Bob's observables,
respectively. We define the Bell correlations with respect to $\scr{A}$ and $\scr{B}$ in $\omega$
as
\begin{equation} \label{eq:1}
  \beta(\omega) = \frac{1}{2} \sup \omega(A_1(B_1+B_2)+A_2(B_1-B_2)),
\end{equation}
where the supremum is taken over all selfadjoint $A_i \in \scr{A}$, $B_j \in \scr{B}$ satisfying
$-\idty \leq A_i \leq \idty$, $-\idty \leq B_j \leq \idty$, for $i,j=1,2$. In other words $A_1,A_2$ and
$B_1,B_2$ are (appropriately bounded) observables measurable by Alice respectively Bob. Of course,
a classically correlated (separable) state, or any other state consistent with a local hidden
variable model \cite{Werner89} satisfies the Bell-CHSH-inequality $\beta(\omega)\leq1$.

Exactly as in the standard case, we can show Cirelson's inequality
\cite{Cirelson,SumWer95,BellRev} bounding the quantum violations of the inequality as
\begin{equation}
  \beta(\omega) \leq \sqrt{2}.
\end{equation}
If the upper bound $\sqrt{2}$ is attained we speak of a \emph{maximal violation} of Bell's
inequality.

It is clear that the maximally entangled state described above does saturate this bound: In the
infinite tensor product construction of $\HH=\Hii$ we only need to take observables $A_i,B_i$ from
the first tensor factor. But we could also have chosen similar observables $A_{i,k},B_{i,k}$
($i=1,2$) for the $k\,^{\rm th}$ qubit pair. Let us denote by
\begin{equation}\label{testop}
  T_k= A_{1,k}(B_{1,k}+B_{2,k})+A_{2,k}(B_{1,k}-B_{2,k})
\end{equation}
the ``test operator'' for the $k\,^{\rm th}$ qubit pair, whose expectation enters the
Bell-CHSH-inequality. Then for a dense set of vectors $\phi\in\HH$, namely for those differing from
the reference vector in only finitely many positions, we get $\langle\phi,T_k\phi\rangle=\sqrt2$
for all sufficiently large $k$. Since the norms $\norm{T_k}$ are uniformly bounded, a simple
$3\varepsilon$-argument shows that $\lim_{k\to\infty}\langle\phi,T_k\phi\rangle=\sqrt2$ for all
$\phi\in\Hii$. By taking mixtures we find
\begin{equation}\label{locnorm}
  \lim_{k\to\infty}\tr(\rho T_k)=\sqrt2
\end{equation}
for all density operators $\rho$ on $\Hii$.

This property is clearly impossible in the finite dimensional case: any product state would violate
it. This clarifies the statement in Section~\ref{sec:von-neumann-algebras}
that $\Hii$ is in no way a tensor product of
Hilbert spaces for Alice and Bob. Of course, we can simply \emph{define} a product state on the
algebra of local operators, and then extend it by the Hahn-Banach Theorem to all operators on
$\B(\Hii)$. However, just as the reference state of infinitely many singlets is a singular state on
$\B(\Hi\otimes\Hi)$, any product state will necessarily be singular on $\B(\Hii)$.

It is interesting that bipartite states with property (\ref{locnorm}) naturally arise in quantum
field theory, with $\A$ and $\B$ the algebras of observables measurable in two causally disjoint
(but tangent) spacetime regions. This is true under axiomatic assumptions on the structure of local
algebras, believed to hold in any free or interacting theory. The only thing that enters is indeed
the structure of the local von Neumann algebras, as shown by the
following Theorem \cite{SumWer87,SumWer88,SumWer95}. Again the
maximally entangled state plays a key role.

\begin{The}{\rm (\cite{SumWer88})} Let $\A,\B\subset\B(\HH)$ be mutually commuting von Neumann algebras acting on a
separable Hilbert space $\HH$. Then the following are equivalent:
\begin{description}
\item[\rm (i)] For some density operator $\rho$, which has no zero eigenvalues, we have $\beta(\rho)=\sqrt2$.
\item[\rm (ii)] For every density operator $\rho$ on $\HH$ we have $\beta(\rho)=\sqrt2$.
\item[\rm (iii)] There is a set $T_k$ of test operators formed from $\A$ and $\B$ such that (\ref{locnorm})
       holds for all density operators $\rho$.
\item[\rm (iv)] There is a unitary isomorphism under which
\begin{eqnarray}
  \HH&=&\Hii\otimes\widetilde\HH  \;,\nonumber\\
  \A&=&\A_1\otimes\widetilde\A    \;,\nonumber\\
  \B&=&\B_1\otimes\widetilde\B    \;,\nonumber
\end{eqnarray}
  $\A_1,\B_1\subset\B(\Hii)$ are the algebras of Theorem~\ref{maxentio}, and
   $\widetilde\A,\widetilde\B\subset\B(\widetilde\HH)$ are other von Neumann algebras.
\end{description}
\end{The}

In other words, the maximal violation of Bell's inequalities for \emph{all} normal states implies
that the bipartite system is precisely the maximal entangled state, plus some additional degrees of
freedom ($\widetilde\A,\widetilde\B$), which do not contribute to the violation of Bell
inequalities.

\subsection{Schmidt decomposition and modular theory}
\label{sec:schm-decomp-modul}

The Schmidt decomposition is a key technique for analyzing bipartite pure states in the standard
framework. It represents an arbitrary vector $\Omega\in\HH_A\otimes\HH_B$ as
\begin{equation}\label{schmitt}
  \Omega=\sum_\alpha c_\alpha \,e_\alpha\otimes f_\beta\;,
\end{equation}
where the $c_\alpha>0$ are positive constants, and $\{e_\alpha\}\subset\HH_A$ and
$\{f_\alpha\}\subset\HH_B$ are orthonormal systems.

Its analog in the context of von Neumann algebras is a highly developed theory with many
applications in quantum field theory and statistical mechanics, known as the \emph{modular theory}
of Tomita and Takesaki \cite{Tak70}. We recommend Chapter 2.5 in \cite{BraRob} for an excellent
exposition, and only outline some ideas and indicate the connection to the Schmidt decomposition.

Throughout this subsection, we will assume that $\A,\B\subset\B(\HH)$ are von Neumann algebras, and
$\Omega\in\HH$ is a unit vector, such that the properties ME 2, ME 3, and ME 7 of Section~\ref{sec:maxent}
hold. As in the case of the usual Schmidt decomposition the essential information is already
contained in the restriction of the given state to the subalgebra $\A$, i.e., by the linear
functional $\omega(A)=\langle\Omega,A\Omega\rangle$. Indeed, the Hilbert space and the cyclic
vector $\Omega$ (cf. ME 7) satisfy precisely the conditions for the GNS-representation, which is
unique up to unitary equivalence.  Moreover, condition ME 2 fixes $\B$ as the commutant algebra.

However, since $\A$ often does not admit a trace, we cannot represent $\omega$ by a density
operator, and therefore we cannot use the spectrum of the density operator to characterize
$\omega$. Surprisingly, it is equilibrium statistical mechanics, which provides the notion to
generalize. In the finite dimensional context, we can consider every density operator as a
canonical equilibrium state, and determine from it the Hamiltonian of the system. This in turn
defines a time evolution. Note that the Hamiltonian is only defined up to a constant, so we cannot
expect to reconstruct the eigenvalues of $H$, but only the spectrum of the Liouville operator
$\sigma\mapsto i[\sigma,H]$, which generates the dynamics on density operators, and has eigenvalues
$i(E_n-E_m)$, when the $E_n$ are the eigenvalues of $H$. The connection between the time evolutions
and equilibrium states makes sense also for von Neumann algebras, and can be seen as the physical
interpretation of modular theory \cite{BraRob}.

We begin the outline of this theory with the anti-linear operator $S$ on $\scr{H}$ by
\begin{equation}\label{modcon}
  S (A\Omega) = A^*\Omega,\quad A \in \scr{A}.
\end{equation}
It turns out to be closable, and we denote its closure by the same letter. As a closed operator $S$
admits a polar decomposition
\begin{equation}
  S = J \Delta^{1/2},
\end{equation}
which defines the anti-unitary \emph{modular conjugation} $J$ and the positive \emph{modular
operator} $\Delta$.

Let us calculate $\Delta$ in the standard situation, where $\scr{H} = \scr{K} \otimes \scr{K}$, and
$\scr{A} = \scr{B}(\scr{K}) \otimes \idty$ respectively $\scr{B} = \idty \otimes \scr{B}(\scr{K})$,
and $\Omega$ is in Schmidt form (\ref{schmitt}). Due to assumption ME 7 (cyclicity), the orthonormal
systems $e_\alpha$ and $f_\alpha$ have to be even complete (i.e., bases). Now consider
(\ref{modcon}) with $A=(\ketbra{e_\beta}{e_\gamma})\otimes\idty$, which becomes
\begin{equation}\label{modcon-1}
  S( c_\gamma e_\beta\otimes f_\gamma)
  = c_\beta e_\gamma\otimes f_\beta,
\end{equation}
from which we readily get
\begin{equation} \label{eq:3}
  \Delta^{1/2} = \rho^{1/2} \otimes \rho^{-1/2},\quad\mbox{and}\quad
  J = F (\Theta \otimes \Theta),
\end{equation}
where $\rho=\sum_\alpha c_\alpha^2 \kb{e_\alpha}$ is the reduced density operator,  $F \phi_1
\otimes \phi_2 = \phi_2 \otimes \phi_1$ is the flip operator and $\Theta$ denotes complex
conjugation in the $e_n$ basis. The time evolution with Hamiltonian $H=-\log\rho + c\idty$, for
which $\omega$ is now the equilibrium state with unit temperature, is then given by ${\cal
E}_t(A)\otimes\idty=\Delta^{it}(A\otimes\idty)\Delta^{-it}$.

In the case of general von Neumann algebras, the spectrum of $\Delta$ need no longer be discrete,
and it can be a general positive, but unbounded selfadjoint operator.
It turns out that $\Delta^{it}$ still defines a time evolution on
the algebra $\A$, the so-called {\em modular evolution}
The equilibrium condition
cannot be written directly in the Gibbs form $\rho\propto\exp(-H)$, since there is no density
matrix any more, but has to be replaced by the so-called KMS-condition, a boundary condition for
the analytic continuation of correlation functions
\cite{BraRob,KMS} which links the modular evolution to the state.

In the standard situation, the eigenvalue $1$ of $\Delta$ plays a special role, because it points
to degeneracies in the Schmidt spectrum. In the extreme case of a maximally entangled state all
$c_\alpha$ are equal, and $\Delta=\idty$ or, equivalently, $S$ is anti-unitary. This characterization
of maximal entanglement carries over to the von Neumann algebra case: $S$ is anti-unitary if and
only if for all $A_1,A_2\in\A$
\begin{eqnarray}
   \langle\Omega,A_1A_2\Omega\rangle
    &=&\langle A_1^*\Omega,A_2\Omega\rangle
     =\langle SA_1\Omega,SA_2^*\Omega\rangle \nonumber\\
    &=&\langle A_2^*\Omega,A_1\Omega\rangle
     =\langle\Omega,A_2A_1\Omega\rangle.\nonumber
\end{eqnarray}
This is precisely the trace property ME 4.

\subsection{Characterization by the EPR-doubles property}
\label{sec:EPR-prop}

In the original EPR-argument it is crucial that certain observables of Alice and Bob are perfectly
correlated, so that Alice can find the values of observables on Bob's side with certainty, without
Bob having to carry out this measurement. An approach to studying such correlations was proposed
recently by Arens and Varadarajan \cite{AV}. The basic idea, stripped of some measure theoretic
overhead, and extended to the more general bipartite systems considered here \cite{WAV}, rests on
the following definition. Let $\A,\B$ be commuting observable algebras and $\omega$ a state on an
algebra containing both $\A$ and $\B$. Then we say that an element $B\in\B$ is an {\it EPR-double}
of $A\in\A$, or that $A$ and $B$ are doubles (of each other) if
\begin{equation}\label{double}
  \omega\bigl((A^*-B^*)(A-B)\bigr)
   =\omega\bigl((A-B)(A^*-B^*)\bigr)
   =0.
\end{equation}
Of course, when $A$ and $B$ are hermitian, the two expressions coincide, and in this case there is
a simple interpretation of equation (\ref{double}). Since $A$ and $B$ commute, we can consider
their joint distribution (measuring the joint spectral resolution of $A$ and $B$). Then $(A-B)^2$ is a
positive quantity, which has vanishing expectation if and only if the joint distribution is
concentrated on the diagonal, i.e., if the measured values coincide with probability one.

Basic properties are summarized in the following Lemma.

\begin{Lemma}\label{doublemma} Let $\omega$ be a state on a C*-algebra
containing commuting subalgebras $\A$ and $\B$. Then
\begin{description}
\item[\rm (i)]
$A$ and $B$ are doubles iff for all $C$ in the ambient observable algebra we have
$\omega(AC)=\omega(BC)$ and $\omega(CA)=\omega(CB)$.

\item[\rm (ii)]
If $A_1,A_2$ have doubles $B_1,B_2$, then $A_1^*, A_1+A_2$, and $A_1A_2$ have doubles
 $B_1^*, B_1+B_2$, and $B_2B_1$, respectively.

\item[\rm (iii)]
When $A$ and $B$ are normal ($AA^*=A^*A$), and doubles of each other, then so are $f(A)$ and
$f(B)$, where $f$ is any continuous complex valued function on the spectrum of $A$ and $B$,
evaluated in the functional calculus.

\item[\rm (iv)]
When $\A$ and $\B$ are von Neumann algebras, and $\omega$ is a normal state, and observables
 $A_n$ with doubles $B_n$ converge in weak*-topology to $A$, then every cluster point of the
 sequence $B_n$ is a double of $A$.

\item[\rm (v)]
Suppose that $\omega$ restricted to $\B$ is faithful (i.e., $\B\ni B\geq0$ and $\omega(B)=0$ imply
 $B=0$). Then every $A\in\A$ admits at most one double.
\end{description}
\end{Lemma}

\begin{proof}
(i) One direction is obvious by setting $C=A^*-B^*$. The other direction follows from the
Schwartz inequality $|\omega(X^*Y)|^2\leq\omega(X^*X)\omega(Y^*Y)$.

The remaining items follow directly from (i). (iii) is obvious from (ii) for polynomials in $A$ and
$A^*$, and extends to continuous functions by taking norm limits on the polynomial approximations
to $f$ provided by the Stone-Weierstra{\ss}\ approximation theorem. For (iv) one has to use the
weak*-continuity of the product in each factor separately (see e.g. \cite[Theorem 1.7.8]{Sak}).
\end{proof}

In the situation we have assumed for modular theory, we can give a detailed characterization of the
elements admitting a double:

\begin{Pro} Suppose $\A$ and $\B=\A'$ are von Neumann algebras on a Hilbert space $\HH$, and the
state $\omega$ is given by a vector $\Omega\in\HH$, which is cyclic for both $\A$ and $\B$. Then
for every $A\in\A$ the following conditions are equivalent:
\begin{description}
\item[\rm (i)]
$A$ has an EPR-double $B\in\B$.

\item[\rm (ii)]
$A$ is in the \emph{centralizer} of the restricted state, i.e.,
   $\omega(AA_1)=\omega(A_1A)$ for all $A_1\in\A$.

\item[\rm (iii)]
$A$ is invariant under the modular evolution
   $\Delta^{it}A\Delta^{-it}=A$ for all $t\in\Rl$.
\end{description}
In this case the double is given by $B=JA^*J$.
\end{Pro}

\begin{proof}
(i)$\Rightarrow$(ii) When $A$ has a double $B$, we get
$\omega(AA_1)=\omega(BA_1)=\omega(A_1B)=\omega(A_1A)$
for all $A_1$ in the ambient observable algebra.

(ii)$\Leftrightarrow$(iii) This is a standard result (see, e.g., \cite[Prop. 15.1.7]{BaumWoll}).

(iii)$\Rightarrow$(i) Since $\Delta^{it}\Omega=\Omega$, (iii) implies $\Delta^{it} A\Omega=A\Omega$, so
$A\Omega$ is an eigenvector for eigenvalue 1 of the unitary $\Delta^{it}$ and $\Delta
A\Omega=A\Omega$. By the same token, $\Delta A^*\Omega=A^*\Omega$. We claim that in that case
$B=JA^*J\in\B$ is a double of $A$ in $\B$: We have $B\Omega=JA^*J\Omega=JA^*\Omega=JSA\Omega=\Delta
A\Omega=A\Omega$ and, similarly, $B^*\Omega=A^*\Omega$. From this (i) follows immediately.

The formula for $B$ was established in the last part of the proof. Uniqueness follows from
Lemma~\ref{doublemma}.
\end{proof}

Two special cases are of interest. On the one hand, in the standard case of a pure bipartite state
we get a complete characterization of the observables which posses a double: they are exactly the
ones commuting with the reduced density operator \cite{AV}. On the other hand, we can ask under
what circumstances {\it all} $A\in\A$ admit a double. Clearly, this is the case when the
centralizer in (ii) of the Proposition is all of $\A$, i.e., if and only if the restricted state is
a trace. Again this characterizes the everybody's maximally entangled states on finite dimensional
algebras, and the unique infinite dimensional one for hyperfinite von Neumann algebras.

\section{The original EPR state}\label{sec:EPR}

In their famous 1935 paper \cite{EPR35} Einstein, Podolsky and Rosen  studied two quantum particles
with perfectly correlated momenta and perfectly anticorrelated positions. It is immediately clear
that such a state does not exist in the standard framework of Hilbert space theory: the difference
of the positions is a self-adjoint operator with purely absolutely continuous spectrum, so whatever
density matrix we choose, the probability distribution of this quantity will have a probability
density with respect to Lebesgue measure, and cannot be concentrated on a single point.
Consequently, the wave function written in \cite{EPR35} is a pretty wild object. Essentially it is
$\Psi(x_1,x_2)=c \delta(x_1-x_2+a)$, with the Dirac delta function, and $c$ a ``normalization
factor'' which must vanish, because the normalization integral for the delta function is undefined,
but infinite if anything.

How could such a profound physical argument be based on such an ill-defined object? The answer is
probably that the authors were completely aware that they were really talking about a limiting
situation of more and more sharply peaked wave functions. We could model them by a sequence of more
and more highly squeezed two mode Gaussian states (cf. Subsection \ref{sec:epr-states-based}), or some
other sequence representation of the delta function. The key point is that the main argument does not
depend on the particular approximating sequence. But then we should also be able to discuss the limiting
situation directly in a rigorous way, and extract precisely what is common to all approximations of 
the EPR state.

\subsection{Definition}
In this section we consider a family of singular states, which describes quite well what Einstein
Podolsky and Rosen may have had in mind. Throughout we assume we are in the usual Hilbert space
${\scr H}=\scr{L}^2(\Rl^2)$ for describing two canonical degrees of freedom, with position and
momentum operators $Q_1,Q_2,P_1,P_2$. The basic observation is that the operators $P_1+P_2$ and
$Q_1-Q_2$ commute as a consequence of the Heisenberg commutation relations. Therefore we can
evaluate in the functional calculus (i.e., using a joint spectral resolution) any function of the
form $g(P_1+P_2,Q_1-Q_2)$, where $g:\Rl^2\to\Cx$ is an arbitrary bounded continuous function. We
define an {\it EPR-state} as any state $\omega$ such that
\begin{equation}\label{EPRdef}
 \omega\Bigl(g(P_1+P_2,Q_1-Q_2)\Bigr)=g(0,a)\;,
\end{equation}
where $a$ is the fixed distance between the particles. Several comments are in order. First of all,
if we take any sequence of vectors to ``approximate'' the EPR wave function (and adjust
normalization on the way), weak*-cluster points of the corresponding sequence of pure states exist
by compactness of the state space, and all these will be EPR states in the sense of our definition.
Secondly, condition~(\ref{EPRdef}) does not fix $\omega$ uniquely. Indeed, different approximating
sequences may lead to different $\omega$. Even for a fixed approximating sequence it is rarely the
case that the expectation values of \emph{all} bounded operators converge, so the sequence will
have many different cluster points. Thirdly, the existence of EPR states  can also be seen more
directly: the algebra of bounded continuous functions on $\Rl^2$ is faithfully represented in
${\scr B}({\scr H})$ (i.e., $g(P_1+P_2,Q_1-Q_2)=0$ only when $g$ is the zero function). On that
algebra the point evaluation at $(0,a)$ is a well defined state, so any Hahn-Banach extension of
this state to all of ${\scr B}({\scr H})$ will be an EPR state \footnote{The reason for defining
EPR-states with respect to \emph{continuous} functions of $P_1+P_2$ and $Q_1-Q_2$ rather than, say,
measurable functions, is that we need faithfulness. The functional calculus is well defined also
for measurable functions, but some functions will evaluate to zero. In particular, for the function
$g(p,x)=1$ for $x=a$ and $p=0$, but $g(p,x)=0$ for all other points, we get $g(P_1+P_2,Q_1-Q_2)=0$,
because the joint spectrum of these operators is purely absolutely continuous. Hence
condition~(\ref{EPRdef}), extended to measurable functions would require the expectation of the
zero operator to be $1$.}.

In our further analysis we will only look at properties which are common to all EPR states, and
which are hence independent of any choice of approximating sequences. The basic technique for
extracting such properties from (\ref{EPRdef}) is to use positivity of $\omega$ in the form of the
Schwartz inequality $|\omega(A^*B)|\leq\omega(A^*A)\omega(B^*B)$. For example, we get
\begin{equation}\label{gextract}
  \omega\Bigl(X\widehat g\Bigr)
  =\omega\Bigl(\widehat g\;X\Bigr)
  =g(0,a)\omega(X)\;,
\end{equation}
where $\widehat g$ is shorthand for $g(P_1+P_2,Q_1-Q_2)$ for some bounded continuous function $g$,
and $X\in{\scr B}({\scr H})$ is an arbitrary bounded operator. This is shown by taking $A=X^*$ and
$B=(\widehat g-g(0,a)\idty)$ (or $A=(\widehat g-g(0,a)\idty)$ and $B=X$) in the Schwartz
inequality.

\subsection{Restriction to the CCR-algebra}\label{sec:EPRCCR}

Next we consider the expectations of \emph{Weyl operators}
\begin{eqnarray}\label{Weylops}
  \weyl(\xi_1,\xi_2,\eta_1,\eta_2)
   &=&{\rm e}^{{\rm i}(\xi_1 P_1+\xi_2
   P_2-\eta_1Q_1-\eta_2Q_2)}
       \nonumber\\
   &=&{\rm e}^{{\rm i}(\vec{\xi}\cdot\vec{P}-\vec{\eta}\cdot\vec{Q})}\;.
\end{eqnarray}
Obviously, if $\xi_1=\xi_2$ and $\eta_1=-\eta_2$, which we will abbreviate as
$(\vec\xi,\vec\eta)\in S$, we have $\weyl(\vec\xi,\vec\eta)=\widehat g$ for a uniformly continuous
$g$, so (\ref{EPRdef}) determines the expectation. Combining it with Equation~(\ref{gextract}) we
get:
\begin{multline}\label{wS}
          \omega\Bigl(\weyl(\vec\xi,\vec\eta)X\Bigr)
      = \omega\Bigl(X\weyl(\vec\xi,\vec\eta)\Bigr)
      =\omega(X)\;,\\
      \mbox{for}\hspace{10pt} (\vec\xi,\vec\eta)\in S\;.
\end{multline}
In particular, the state is invariant under all phase space translations by vectors in $S$.

This is already sufficient to conclude that the state is purely singular, i.e., that $\omega(K)=0$
for every compact operator, and in particular for all finite dimensional projections. An even
stronger statement is that the restrictions to Alice's and Bob's subsystem are purely singular.

\begin{Lemma} For any EPR state, and any compact operator $K$, $\omega(K\otimes\idty)=0$.
\end{Lemma}

\begin{proof}Indeed the restricted state is invariant under {\it all} phase space translations, since we
can extend $\weyl(\xi,\eta)$ to a Weyl operator of the total system, i.e.,
$\weyl'(\xi,\eta)=\weyl(\xi,\xi,\eta,-\eta)\cong\weyl(\xi,\eta)\otimes\weyl(\xi,-\eta)$, with
$(\xi,\xi,\eta,-\eta)\in S$, and
\begin{eqnarray}\label{EPRAinv}
 &&\omega\bigl((\weyl(\xi,\eta)A\weyl(\xi,\eta)^*)\otimes\idty)\\
 &&\qquad=\omega\bigl(\weyl'(\xi,\eta)(A\otimes\idty)\weyl'(\xi,\eta)^*))\;.\nonumber
\end{eqnarray}
Now consider a unit vector $\chi$ with bounded support in position space, and let
$K=\ketbra\chi\chi$ be the corresponding one-dimensional projection. Then sufficiently widely space
translates $\weyl(n\xi_0,0)\chi$ are orthogonal, and hence, for all $N$, the operator
$K_N=\sum_{n=1}^N \weyl(n\xi_0,0)K\weyl(n\xi_0,0^*)$ is bounded by $\idty$. Hence
$N\omega(K)=\omega(K_N)\leq\omega(\idty)=1$, and $\omega(K)=0$. Since vectors of compact support
are norm dense in Hilbert space, the conclusion holds for arbitrary
\end{proof}

 For other Weyl operators we get the expectations from the Weyl
commutation relations
\begin{multline}\label{weylrel}
  \weyl(\vec\xi,\vec\eta)\weyl(\vec\xi\,',\vec\eta\,')
  ={\rm e}^{{\rm i}\frac{\sigma}{2}} \ \weyl(\vec\xi+\vec\xi\,',\vec\eta+\vec\eta\,')\;,\\
\mbox{with}\hspace{10pt} \sigma=\vec\xi\cdot\vec\eta\,'-\vec\xi\,'\cdot\vec\eta \;
.
\end{multline}
This is just a form of the Heisenberg commutation relations. Now $S$ is a
so-called \emph{maximal isotropic} subspace of phase space, which is to say that the commutation
phase $\sigma$ vanishes for $(\vec\xi,\vec\eta),(\vec\xi\,',\vec\eta\,')\in S$, and no subspace of
phase space strictly including $S$ has the same property.

For a point $(\vec\xi,\vec\eta)$ in phase space, which does not
belong to $S$, we can find some vector $(\vec\xi\,',\vec\eta\,')\in
S$ such that the commutation phase ${\rm e}^{{\rm i}\sigma}\not=1$
is non trivial. Combining the Weyl relations (\ref{weylrel}) with the invariance
(\ref{wS}) gives $\omega(\weyl(\vec\xi,\vec\eta))=
\omega(\weyl(\vec\xi\,',\vec\eta\,')\weyl(\vec\xi,\vec\eta))=
{\rm e}^{{\rm i}\sigma}\omega(\weyl(\vec\xi,\vec\eta)\weyl(\vec\xi\,',\vec\eta\,'))
={\rm e}^{{\rm i}\sigma}\omega(\weyl(\vec\xi,\vec\eta))$ which
implies that the expectation values
\begin{equation}\label{wNotS}
  \omega\Bigl(\weyl(\vec\xi,\vec\eta)\Bigr)=0\qquad \mbox{for }(\vec\xi,\vec\eta)\notin S\;
\end{equation}
must vanish.
With equations (\ref{wS},\ref{wNotS}) we have a complete characterization of the state $\omega$
restricted to the ``CCR-algebra'', which is just the C*-algebra generated by the Weyl operators.
Since this is a well-studied object, one might make these equations the starting point of an
investigation of EPR states. However, one can see that (\ref{EPRdef}) is strictly stronger:
there are states which
look like $\omega$ on the CCR-algebra, but which give an expectation in (\ref{EPRdef})
corresponding to a limit of states going to infinity instead of going to zero.

\subsection{EPR-correlations}\label{sec:EPRcor}

How about the correlation property, which is so important in the EPR-argument? The best way to show
this is the `double' formalism of Section~\ref{sec:EPR-prop} in which we denote by $\alal$ the norm
closed subalgebra of operators on $\scr{L}^2(\Rl)$ generated by all operators of the form
$f(\xi P+\eta Q)$, where $f:\Rl\to\Cx$ is an arbitrary {\it uniformly continuous} function evaluated in
the functional calculus on a real linear combination $\xi P+ \eta Q$ of position and momentum
\footnote{The same type of operators, although motivated by a different argument
already appears in \cite{ClHa01}}.
This
algebra is fairly large: it contains many observables of interest, in particular all Weyl operators
and all compact operators. It is closed under phase space translations, and these act continuously
in the sense that, for $Z\in\alal$, $\norm{\weyl({\xi,\eta})Z\weyl({\xi,\eta})^*-Z}\to0$ as
$(\xi,\eta)\to0$ \footnote{This continuity is crucial in the
correspondence theory set out in \cite{QHA2}. We where not able to prove the
analogue of Theorem \ref{double-thm} by only assuming this
continuity.}.

\begin{The}\label{double-thm}
All operators of the form {\rm $Z\otimes\idty$} with $Z\in\alal$ have doubles in the
sense of equation {\rm (\ref{double})}. Moreover, the double of {\rm $Z\otimes\idty$} is {\rm $\idty\otimes Z^T$},
where $Z^T$ denotes the transpose (adjoint followed by complex conjugation) in the position
representation.
\end{The}

\begin{proof} We only have to show that for $f(\xi P+\eta Q)\otimes\idty=f(\xi P_1+ \eta Q_1)$
we get the double $\idty\otimes f(-\xi P+\eta Q)=f(-\xi P_2+ \eta Q_2)$,
when $f,\xi$, and $\eta$ are as in the definition of $\alal$. By
the general properties of the double construction this will then automatically extend to operator
products and norm limits.

Fix $\varepsilon>0$.  Since $f$ is uniformly continuous, there is some $\delta>0$ such that
$\abs{f(x)-f(y)}\leq\varepsilon$ whenever $\abs{x-y}\leq\delta$. Now pick a continuous function
$h:\Rl\to[0,1]\subset\Rl$ such that $h(0)=1$, $h(t)=0$ for $\abs t>\delta$. We consider the
operator
\begin{eqnarray}
  M&=&(f(\xi P_1+\eta Q_1)-f(-\xi P_2+\eta Q_2))\times \nonumber\\
       &&\qquad\times h(\xi (P_1+P_2)+\eta (Q_1-Q_2))\nonumber\\
    &=&F(\xi P_1+\eta Q_1,-\xi P_2+\eta Q_2)\nonumber
\end{eqnarray}
where $F(x,y)=(f(x)-f(y))h(x-y)$ and this function is evaluated in the functional calculus of the
commuting selfadjoint operators $(\xi P_1+\eta Q_1)$ and $(-\xi P_2+\eta Q_2)$.
But the real valued function $F$
satisfies $\abs{F(x,y)}\leq\varepsilon$ for all $(x,y)$: when $\abs{x-y}>\delta$ the $h$-factor
vanishes, and on the strip $\abs{x-y}\leq\delta$ we have $\abs{f(x)-f(y)}\leq\varepsilon$.
Therefore $\norm M\leq\varepsilon$. Let $X$ be an arbitrary operator. Then
\begin{eqnarray}\label{ae}
  &&\big|\omega\bigl(\bigl[f(\xi P_1+\eta Q_1)-f(-\xi P_2+\eta Q_2)\bigr]X\bigr)\big|\nonumber\\
    &&\qquad=\abs{\omega\bigl(MX \bigr)}
    \leq \norm M\,\norm X
    \leq \varepsilon \norm X\;.\nonumber
\end{eqnarray}
Here we have added a factor $h(\xi(P_1+P_2)+\eta(Q_1-Q_2))$
at the second equality sign, which we may because of
(\ref{gextract}), and because $h$ is a function of the appropriate operators, which is $=1$ at the
origin. Since this estimate holds for any $\varepsilon$, we conclude that the first relation in
Lemma~\ref{doublemma}.1 holds. The argument for the second relation is completely analogous.
\end{proof}

\subsection{Infinite one-shot entanglement}\label{sec:EPR1shot}

In order to show that the EPR state is indeed highly entangled, let us verify that it contains
infinite one-shot entanglement in the sense forbidden by Theorem~\ref{nogo}. The local operations
needed to extract a $d$-dimensional system will be simply the restriction to a subalgebra. In other
words, we will construct subalgebras $\A_d\subset\A$ and $\B_d\subset\B$ such that the state
$\omega$ restricted to $\A_d\otimes\B_d$ will be a maximally entangled pure state of
$d$-dimensional systems.

The matrix algebras $\A_d,\B_d$ are best seen to be generated by Weyl operators, satisfying a
discrete version of the canonical commutation relations (\ref{weylrel}), with the addition
operation on the right hand side replaced by the addition in a finite group. Let $\Ir_d$ denote the
cyclic group of integers modulo $d$. With the canonical basis $\ket{k,\ell},\ k,\ell\in\Ir_d$ we
introduce the Weyl operators
\begin{eqnarray}\label{Weylito}
  &&\Weyl(n_1,m_1,n_2,m_2)\ket{k,\ell}    \\
  &&\qquad=\zeta^{n_1(k-m_1)+n_2(\ell-m_2)}\ket{k-m_1,\ell-m_2},\nonumber
\end{eqnarray}
where $\zeta=\exp(2\pi i/d)$ is the $d\,$th root of unity. These are a basis of the vector space
$\B(\Cx^d\otimes\Cx^d)$, which shows that this algebra is generated by the four unitaries
$u_1=\Weyl(1,0,0,0), v_1=\Weyl(0,1,0,0), u_2=\Weyl(0,0,1,0)$ and $v_2=\Weyl(0,0,0,1)$. They are
defined algebraically by the relations $v_ku_k=\zeta u_kv_k, k=1,2$, and
$u_1^d=u_2^d=v_1^d=v_2^d=\idty$.
The one dimensional projection onto the standard maximally
entangled vector $\Omega=d^{-1/2}\sum_k\ket{kk}$ can be expressed in the basis (\ref{Weylito}) as
\begin{eqnarray}\label{Weylitop}
  \kb\Omega
    &=&\frac1{d^2}\sum_{n,m}\Weyl(n,m,-n,m)     \nonumber\\
    &=&\frac1{d^2}\sum_{n,m}(u_1u_2^{-1})^n(v_1v_2)^m,
\end{eqnarray}
which will be useful for computing fidelity.

In order to define the subalgebras extracting the desired entanglement we first define operators
$U_1,V_1$ in Alice's subalgebra and $U_2,V_2$ in Bob's, which satisfy the above relations and hence
generate two copies of the $d\times d$ matrices. It is easy to satisfy the commutation relations
$V_kU_k=\zeta U_kV_k$, by taking appropriate Weyl operators, say
\begin{equation}\label{weyltil}
 \widetilde U_1={\rm e}^{iQ_1},\
 \widetilde U_2={\rm e}^{i(Q_2-a)},
 \quad\mbox{and}\quad
 \widetilde V_k={\rm e}^{i\xi P_k}
\end{equation}
with $\xi=2\pi/d$. The tilde indicates that these are not quite the operators yet we are looking
for, because they do not satisfy the periodicity relations: $\widetilde
U_1^d=\exp(idQ_1)\neq\idty$, and similarly for $U_2^d$ and $\widetilde V_k$.
We will denote by $\widetilde {\cal A}$ the
C*-algebra, generated by the operators $\widetilde U_1,\widetilde
V_1$  (\ref{weyltil}).
The algebra $\widetilde {\cal B}$ is constructed analogously. Then  by virtue of the commutation
relations $\widetilde U_1^d$ and $\widetilde V_1^d$ commute with all other elements of
$\widetilde{\cal A}$,
i.e., they belong to the \emph{center}
${\cal C}_A\subset \widetilde{\cal A}$, which represents the classical variables of the system.
In the same manner, $\widetilde U_2^d$ and $\widetilde V_2^d$
generate the center ${\cal C}_B$ of Bob's algebra $\widetilde {\cal
B}$
\footnote{The C*-algebra $\widetilde {\cal A}$ is isomorphic to the continuous sections in an
C*-algebra bundle over the torus, where each fiber is a copy of the algebra ${\cal A}_d$. Such a
bundle is called {\it trivial}, if it is isomorphic to  the tensor product ${\cal A}_d\otimes{\cal
C}_A$. This would directly give us the desired subalgebra ${\cal A}_d$ as a subalgebra of $\cal A$.
However, this is bundle is not trivial \cite{BraEll92,HoegSkjel81}.  In order to ``trivialize'' the
bundle, we are therefore forced to go beyond norm continuous operations, which respect the
continuity of bundle sections.  Instead we have to go to the measurable functional calculus, and
introduce an operation on the fibers, which depends discontionuously on the base point, through the
introduction of a branch cut.}.

If we take any continuous function (in the functional calculus) of a
hermitian or unitary element of ${\cal C}_A$,
it will still be in ${\cal C}_A$. If we take a measurable
(possibly discontinuous) function the result may fail to be in ${\cal C}_A$, but it still commutes
with all elements of $\widetilde {\cal A}$ (and analogously for Bob's algebras).
In particular, we construct the operators
\begin{equation}\label{d-roots}
  \widehat U_k=\bigl(\widetilde U_k^d\bigr)^{1/d},
\end{equation}
where the $d$\/th root of numbers on the unit circle is taken with a branch cut on the negative
real axis. This branch cut makes the function discontinuous, and also makes this odd-looking
combination very different from $\widetilde U_k$. We now define $\widehat V_k$ analogously, and set
\begin{equation}\label{weylnotil}
 U_k=\widehat U_k^{-1}\widetilde U_k
 \quad\mbox{and}\quad
 V_k=\widehat V_k^{-1}\widetilde V_k
\end{equation}
for $k=1,2$. Then since $\widehat U_k,\widehat V_k$ commute with
$\widetilde{\cal A}\otimes\widetilde{\cal B}$, the commutation
relations $V_kU_k=\zeta U_kV_k$ still hold, but in addition we have $U_k^d=\idty$, because
$\widehat U_k^d=\widetilde U_k^d$. It remains to show that on the finite dimensional algebras
generated by these operators, the given state is a maximally entangled pure state. We will verify
this by computing the fidelity, i.e., the expectation of the projection (\ref{Weylitop}):
\begin{equation}\label{weylfidel}
  \omega\left(\frac1{d^2}\sum_{n,m}(U_1U_2^{-1})^n(V_1V_2)^m\right)=1.
\end{equation}

\noindent{\it Proof of this equation}.\/ We have shown in Section~\ref{sec:EPRcor} that $\widetilde
U_1$ and $\widetilde U_2$ are EPR-doubles. This property transfers to arbitrary continuous
functions of $\widetilde U_1$ and $\widetilde U_2$ by Lemma~\ref{doublemma} and uniform
approximation of continuous functions by polynomials. However, because the state $\omega$ is not
normal, it does {\it not} transfer automatically to the measurable functional calculus and hence
not automatically to $\widehat U_1$ and $\widehat U_2$. We claim that this is true nonetheless.

Denote by $r_d(z)=z^{1/d}$ the $d$th root function with the branch cut as described, and let
$f_\epsilon$ be a continuous function from the unit circle to the unit interval $[0,1]$ such that
$f_\epsilon(z)=1$ except for $z$ in an $\epsilon$-neighborhood of $z=-1$ in arclength, and such
that $f_\epsilon(-1)=0$. Then the function $z\mapsto f_\epsilon(z)r_d(z)$ is continuous. Then,
since $\widetilde U_1^d$ and $\widetilde U_2^d$ are doubles, so are $f_\epsilon(\widetilde U_1^d)$,
$f_\epsilon(\widetilde U_1^d)\widehat U_1$ and their counterparts. Note that both of these commute
with all other operators involved. Hence (using the notation $|X|^2=X^*X$ or $|X|^2=XX^*$, which
coincide in this case)
\begin{eqnarray}\label{omff1}
 &&\omega\Bigl(f_\epsilon(\widetilde U_1^d)^2|\widehat U_1-\widehat U_2|^2\Bigr)
 \nonumber\\
  &&\qquad=\omega\Bigl(\bigl|f_\epsilon(\widetilde U_1^d)\widehat U_1
                -f_\epsilon(\widetilde U_2^d)\widehat U_2\bigr|\Bigr)=0,
\end{eqnarray}
where the first equality holds by expanding the modulus square, and applying the double property of
$f_\epsilon(\widetilde U_1^d)$ where appropriate. On the other hand, we have
\begin{eqnarray}\label{omff2}
 &&\omega\Bigl(\bigl(\idty-f_\epsilon(\widetilde U_1^d)^2\bigr)
           |\widehat U_1-\widehat U_2|^2\Bigr)
 \nonumber\\&&\qquad
   \leq4\omega\bigl(\idty-f_\epsilon(\widetilde U_1^d)^2\bigr)
   \leq4 \frac\epsilon\pi,
\end{eqnarray}
because $\norm{\widehat U_1-\widehat U_2}\leq2$, and $0\leq f_\epsilon(\widetilde U_1^d)\leq\idty$.
For the estimate we used that $f_\epsilon(z)^2$ for all $z$ on the unit circle except a section of
relative size $2\epsilon/(2\pi)$, and that the probability distribution for the spectrum of
$\widetilde U_1^d$ is uniform, because the expectation of all powers $(\widetilde U_1^d)=\exp(ind
Q_1)$ vanishes.

Adding (\ref{omff1}) and (\ref{omff2}) we find that
 $\omega\bigl(|\widehat U_1-\widehat U_2|^2\bigr)\leq4\epsilon/\pi$ for every $\epsilon$, and hence
that $\widehat U_1$ and $\widehat U_2$ are EPR doubles as claimed. The proof that $\widehat V_1$
and $\widehat V_2^*$ are likewise doubles (just as $\widetilde V_1$ and $\widetilde V_2^*$) is
entirely analogous. Hence $U_1$ and $U_2$ as well as $V_1$ and $V_2$ are also doubles. Applying
this property in the fidelity expression (\ref{weylfidel}) we find that every term has expectation
one, so that with the prefactor $d^{-2}$ the $d^2$ terms add up to one as claimed.
\hfill$\square\quad$

\subsection{EPR states based on two mode Gaussians}
\label{sec:epr-states-based}

In this section we will deviate from the announcement that we intended to study only such
properties of EPR states which follow from the definition alone, and are hence common to all EPR
states. The reason is that there is one particular family, which has a lot of additional symmetry,
and hence more operators admitting doubles, than general EPR states. Moreover, it is very well
known. In fact, most people working in quantum optics probably have a very concrete picture of the
EPR state, or rather of an approximation to this state: since Gaussian states play a prominent role
in the description of lasers, it is natural to consider a Gaussian wave function of the form
\begin{align}
  \Psi_\lambda(x_1,x_2) =&  \frac{1}{\sqrt{\pi}} \exp\left(-\frac{1-\lambda}{4(1+\lambda)}(q_1-q_2)^2 \right. \notag \\
  & \phantom{\frac{1}{\sqrt{\pi}} \exp\left(\right.}\left. - \frac{1+\lambda}{4(1-\lambda)}(q_1+q_2)^2\right)
  \label{NoPax} \\
         \Psi_\lambda \label{NoPaH}
      =&\sqrt{1-\lambda^2}\;\sum_{n=0}^\infty \lambda^n\,
                                 {\bf e}_n\otimes {\bf e}_n\;,
\end{align}
where ${\bf e}_n$ denotes the eigenbasis of the harmonic oscillators $H_i=(P_i^2+Q_i^2)/2$
$(i=1,2)$. This state is also known as the NOPA state, and the parameter $\lambda\in[0,1)$ is
related to the so-called squeezing parameter $r$ by $\lambda = \tanh(r)$. Values around $r=5$ are considered a 
good experimental achievement \cite{KorLeu}. Of course, we are interested in the limit $r\to\infty$, or
$\lambda\to1$.

The $\lambda$-dependence of the wave function can also be written as
\begin{equation}\label{scaleSqueeze}
  \Psi_\lambda(x_1,x_2)
    =\Psi_0(x_1\cosh\eta+x_2\sinh\eta, -x_1\sinh\eta+x_2\cosh\eta)\;,
\end{equation}
where the hyperbolic angle $\eta$ is $r/2$. It is easy to see that for {\it any} wave function
$\Psi_0$ the probability distributions of both $Q_1-Q_2$ and $P_1+P_2$ scale to a point measures at
zero. Hence any cluster point of the associated sequence of states
$\omega_\lambda(X)=\langle\Psi_\lambda,A\Psi_\lambda\rangle$ is an EPR state in the sense of our
definition (with shift parameter $a=0$). Note, however, that the family itself does not converge to 
any state: it is easy to construct observables $X$ for which the expectation $\omega_\lambda(X)$
remains oscillating between $0$ and $1$ as $\lambda\to1$. Here, as in the general case, a single
state can only be obtained by going to a finest subsequence (or by taking the limit along an
ultrafilter).

The virtue of the particular family (\ref{NoPaH}) is that it has especially high symmetry: it is
immediately clear that
\begin{equation}\label{Hisdoubled}
  \Bigl((f(H_1)-f(H_2)\Bigr)\Psi_\lambda=0
\end{equation}
for all $\lambda$, and for all bounded functions $f:\Nl\to\Cx$ of the oscillator Hamiltonians
$H_1,H_2$. This implies that $f(H_1)$  and $f(H_2)$ are doubles with respect to the state
$\omega_\lambda$ for each $\lambda$. Clearly, this property remains valid in the limit along any
subsequence, so all EPR-states obtained as cluster points of the sequence $\omega_\lambda$ also
have $f(H_1)$ in their algebra of doubles. Consequently, the unitaries $U_k(t)=\exp(itH_k)$ are
also doubles of each other, and the limiting states are invariant under the time evolution
$U_{12}(t)=U_1(t)\otimes U_2(-t)$. This is certainly suggestive, because oscillator time evolutions
have an interpretation as linear symplectic transformations on phase space: $Q_k\mapsto Q_k\cos
t\pm P_k\sin t$ and $P_k\mapsto \mp Q_k\sin t+ P_k\cos t$, where the upper sign holds for $k=1$ and
the lower for $k=2$. The subspace $S$ from Section~\ref{sec:EPRcor} is invariant under such rotations,
and one readily verifies that the time evolution $U_{12}(t)$ takes EPR states into EPR states. This 
certainly implies that by averaging we can generate EPR states invariant under this evolution, and
we have clearly just constructed a family with this invariance.

As $\lambda\to1$, the Schmidt spectrum in (\ref{NoPaH}) becomes ``flatter'', which suggests that
exchanging some labels $n$ should also define a unitary with double. Let $p:\Nl\to\Nl$ denote an
injective (i.e., one-to-one but not necessarily onto map). Then we define an isometry $V_p$ by
\begin{equation}\label{Vperm}
  V_p\,{\bf e}_n={\bf e}_{p(n)}
\end{equation}
with adjoint
\begin{equation}\label{Vpermstar}
  V_p^*\,{\bf e}_n
     =\left\{\begin{array}{ll}
            {\bf e}_{p^{-1}(n)}&\mbox{if\ }n\in p(\Nl)\\
            0&\mbox{if\ }n\notin p(\Nl)\end{array}\right.
\end{equation}
Let us assume that $p$ has {\it finite distance}, i.e., there is a constant $\ell$ such that
$\abs{p(n)-n}\leq\ell$ for all $n\in\Nl$. We claim that in this case $V_p\otimes\idty$ and
$\idty\otimes V_p^*$ are doubles in all EPR states constructed from the sequence (\ref{NoPaH}). We
show this by verifying that the condition holds approximately already for finite $\lambda$.
Consider the vector
\begin{eqnarray}\label{aa}
 \Delta_\lambda&=&\bigl(V_p\otimes\idty-\idty\otimes V_p^*\bigr)\Psi_\lambda \\
  &=&\sqrt{1-\lambda^2}\;\sum_{n=0}^\infty (\lambda^n-\lambda^{p(n)})\,
                                 {\bf e}_{p(n)}\otimes {\bf e}_n\;,\nonumber
\end{eqnarray}
where in the second summand we changed the summation index from $n$ to $p(n)$, automatically
omitting all terms annihilated by $V_p^*$ according to (\ref{Vpermstar}). Since this is a sum of
orthogonal vectors, we can readily estimate the norm by writing
$(\lambda^n-\lambda^{p(n)})=\lambda^n(1-\lambda^{p(n)-n})$:
\begin{equation}
  \norm{\Delta_\lambda}^2\leq \max_n \abs{1-\lambda^{p(n)-n}}^2
    \leq\abs{1-\lambda^{-\ell}}^2
       \;,
\end{equation}
which goes to zero as $\lambda\to1$. Therefore
\begin{equation}
  \omega_\lambda\Bigr(X\bigl(V_p\otimes\idty-\idty\otimes V_p^*\bigr)\Bigr)
   =\langle\Psi_\lambda, X\Delta_\lambda\rangle
   \to0
\end{equation}
as $\lambda\to1$. Hence $V_p\otimes\idty$ and $\idty\otimes V_p^*$ are doubles in any state defined
by a limit of $\omega_\lambda$ along a subsequence, as claimed.

$V_p$ is an isometry but not necessarily unitary. But it is effectively unitary under an EPR state:
Since $V_p$ is in the centralizer, we must have
$\omega\bigl((\idty-V_pV_p^*)\otimes\idty\bigr)=\omega\bigl((\idty-V_p^*V_p)\otimes\idty\bigr)=0$,
although this operator is non-zero. This is in keeping with the general properties of EPR states,
whose restrictions must be purely singular. In fact,  $(\idty-V_pV_p^*)$ is the projection onto
those eigenstates ${\bf e}_n$ for which $n\notin p(\Nl)$, and this set is finite: it has at most
$\ell$ elements\footnote{For any $N>\ell$, consider the set $\{1,\ldots,N\}$. This has to
contain at least the images of  $\{1,\ldots,N-\ell\}$, hence it can contain at most
$\ell$ elements not in $p(\Nl)$.}.

It is interesting to note what happens if one tries to relax the finite distance condition. An
extreme case would be the two isometries $V_{\rm even}\,{\bf e}_n={\bf e}_{2n}$ and $V_{\rm
odd}\,{\bf e}_n={\bf e}_{2n+1}$. These cannot have doubles in {\it any} state, because  the
restriction $\omega_A$ of the state to the first factor would then have to satisfy
 $1=\omega_A(V_{\rm even}V_{\rm even}^*+V_{\rm odd}V_{\rm odd}^*)
   =\omega_A(V_{\rm even}^*V_{\rm even}+V_{\rm odd}^*V_{\rm odd})
   =\omega_A(\idty+\idty)=2$.
On the other hand, the norm of $\Delta_\lambda$ no longer goes to zero, and we get
$\norm{\Delta_\lambda}^2\to1/6$ instead.

To get infinite one-shot entanglement is easier than in the case of general EPR states: we can
simply combine $d$ periodic multiplication operators with $d$-periodic permutation operators to
construct a finite Weyl-system of doubles\footnote{This is probably what the authors of
\cite{Zeil02} are trying to say.}. 
In fact there is a very
quick way to get high fidelity entangled pure states even for $\lambda<1$ (see \cite{NeuDipl} for an
application to Bell inequality violations). Consider the unitary operator
$U_d:\HH\to\HH\otimes\Cx^d$ given by
\begin{equation}
  U_d\,{\bf e}_{dk+r}=({\bf e}_k\otimes{\bf e}^{(d)}_r)\;,
\end{equation}
for $k=0,1,\ldots$ and $r=0,1,\ldots,d-1$. Then
\begin{equation}\label{factorNOPA}
 (U_d\otimes U_d)\Psi_\lambda=\Psi_{\lambda^d}\otimes\Psi^{(d)}_\lambda
\end{equation}
with a $\lambda$-dependent normalized vector $\Psi^{(d)}_\lambda\in\Cx^d\otimes\Cx^d$ proportional
to
\begin{equation}\label{NOPAd}
  \Psi^{(d)}_\lambda \propto \sum_{r=1}^d \lambda^r\, {\bf e}_r^{(d)}\otimes {\bf e}_r^{(d)} \;.
\end{equation}
Note that the infinite dimensional factor on the right hand side of (\ref{factorNOPA}) is again a
state of the form (\ref{NoPaH}), however, a less entangled one with parameter
$\lambda'=\lambda^d<\lambda$. The second factor, i.e., (\ref{factorNOPA}) becomes maximally
entangled in the limit $\lambda\to 1$. Therefore the unitary $(U_d\otimes U_d)$ splits both Alice's
and Bob's subsystem, so that the total system is split exactly into a less entangled version of
itself and a pure, nearly maximally entangled $d$-dimensional pair.  The local operation extracting
entanglement from this state is to discard the infinite dimensional parts. Seen in one of the limit
states of the family $\omega_\lambda$ this is maximally entangled, so equation (\ref{infentR}) is
satisfied with $\epsilon=0$. Moreover, since the remaining system is of exactly the same type, the
process can be repeated arbitrarily often.

\subsection{Counterintuitive properties of the restricted states}

Basically, subsection \ref{sec:EPRcor} shows that the EPR states constructed here do satisfy the
requirements of the EPR argument. However, Einstein, Podolsky and Rosen do not consider the
measurement of suitable periodic functions of $Q_k$ or $P_k$ but measurements of these quantities
themselves \cite{EPR35}: What do EPR states have to say about these?

Unfortunately, the ``values of momentum'' found by Alice or Bob are not quite what we usually mean
by ``values'': they are infinite with probability 1. To see this, recall the remark after eq.
(\ref{wS}) that EPR states are invariant with respect to phase space translations with
$\weyl(\vec\xi,\vec\eta)$ with $(\vec\xi,\vec\eta)\in S$. Hence
\begin{eqnarray}\label{restinv}
&&\hspace{-20pt}\omega\bigl(\weyl(\xi_1,0,\eta_1,0)(A\otimes\idty)\weyl(\xi_1,0,\eta_1,0)^*\bigr)
\nonumber\\
&=&\omega\bigl(\weyl(\xi_1,\xi_1,\eta_1,-\eta_1)(A\otimes\idty)\weyl(\xi_1,\xi_1,\eta_1,-\eta_1)^*\bigr)
\nonumber\\
&=&\omega(A\otimes\idty).
\end{eqnarray}
That is, the reduced state is invariant under all phase space translations. Now suppose that for
some continuous function $f$ with compact support we have $\omega(f(Q_1))=\epsilon\neq0$. Then we
could add many (say $N$) sufficiently widely spaced translates of $f$ to get an operator
$F=\sum_i^Nf(Q_1+x_i\idty)$ with $\norm F\leq\norm f$ and $|N\epsilon|=|\omega(F)|\leq\norm f$,
which implies $\epsilon=0$. Hence for every function with compact support we must have
$\omega(f(Q_1))=0$. Note that this is possible only for singular states, since we can easily
construct a sequence of compactly supported function increasing to the identity,
whose $\omega$ expectations are all zero, hence fail to converge to $1$.

In spite of being infinite, the ``measured values'' of Alice and Bob are perfectly correlated,
which means that we have to distinguish different kinds if infinity. Such ``kinds of infinity'' are
the subject of the topological theory of \emph{compactifications} \cite{Chandler,QHA2}. The basic
idea is very simple: consider some C*-algebra of bounded functions on the real line. Then the
evaluations of the functions at a point, i.e., the functionals $x\mapsto f(x)$, are pure states on
such an algebra, but {\'{}}compactness of the state space together with the Kre\u{\i}n-Milman Theorem
\cite{Alfsen} dictates that there are many more pure states. These additional pure states are
interpreted as the points at infinity associated with the given observable algebra. The set of all
pure states is called the Gel'fand spectrum of the commutative C*-algebra\cite[Sec.2.3.5]{BraRob},
and the algebra is known to be isomorphic to the algebra of continuous functions on this compact
space. For the algebra of all bounded function the additional pure states are called free
ultrafilters, for the algebra of all continuous bounded functions we get the points of the Stone-\v
Cech-compactification, and for the algebra of uniformly continuous functions we get a still coarser
notion of points at infinity. According to Section~\ref{sec:EPRcor} these are the measured values,
which will be perfectly correlated between Alice's and Bob's positions or momenta. It is not
possible to exhibit any such value, because proving their mere existence already requires an
argument based on the Axiom of Choice.

So do we have to be content with the statement that the measured values lie ``out there on the
infinite ranges, where the free ultrafilters roam?'' Section~\ref{sec:EPR1shot} shows that for many
concrete problems, involving not too large observable algebras, we can use the perfect correlation
property quite well. A smaller algebra of observables means that many points of Gel'fand spectrum
become identified, and some of these coarser points may have a direct physical interpretation. So the
moral is not so much that compactification points at infinity are wild, pathological objects, but
that they describe the way a sequence can go to infinity in the finest possible detail, which is
just much finer that we usually want to know. The EPR correlation property holds even for such wild
``measured values''.


\begin{thebibliography}{10}

\bibitem{EisSimPle01}
J.~Eisert, C.~Simon and M.B. Plenio.
\newblock \emph{On the quantification of entanglement in infinite-dimensional
  quantum systems}.
\newblock quant-ph/0112064 (2001).

\bibitem{BSST}
C.~H. Bennett, P.~W. Shor, J.~A. Smolin and A.~V. Thapliyal.
\newblock \emph{Entanglement-assisted capacity of a quantum channel and the
  reverse {Shannon} theorem}.
\newblock quant-ph/0106052 (2001).

\bibitem{SumWer87}
S.J. Summers and R.F.Werner.
\newblock \emph{Maximal violation of {Bell's} inequality is generic in quantum
  field theory}.
\newblock Commun. Math. Phys. \textbf{110}, 247--259 (1987).

\bibitem{SumWer88}
S.J. Summers and R.F.Werner.
\newblock \emph{Maximal violation of {Bell's} inequalities for algebras of
  observables in tangent spacetime regions}.
\newblock Ann. Inst. H. Poincar{\'e} \textbf{A 49}, 215--243 (1988).

\bibitem{EPR35}
A.~Einstein, B.~Podolsky and N.~Rosen.
\newblock \emph{Can quantum-mechanical description of physical reality be
  considered complete?}
\newblock Phys. Rev \textbf{47}, 777--780 (1935).

\bibitem{CliftHalv98}
R.~Clifton and H.~Halvorson.
\newblock \emph{Maximal beable subalgebras of quantum-mechanical observables}.
\newblock Int. J. Theor. Phys. \textbf{38}, 2441--2484 (1999).

\bibitem{CliftHalv99}
R.~Clifton and H.~Halvorson.
\newblock \emph{Bipartite mixied states of infinite dimensional systems are
  generically nonseparable}.
\newblock Phys. Rev. A \textbf{61}, 012108 (2000).

\bibitem{ClHa01}
R.~Clifton and H.~Halvorson.
\newblock \emph{Reconsidering {Bohr's} reply to {EPR}}.
\newblock quant-ph/0110107 (2001).

\bibitem{HorCirLew}
P.~Horodecki, J.I. Cirac and M.~Lewenstein.
\newblock \emph{Bound entanglement for continuous variables is a rare
  phenomenon}.
\newblock quant-ph/0103076 (2001).

\bibitem{Rain00}
E.~Rains.
\newblock \emph{A semidefinite program for distillable entanglement}.
\newblock quant-ph/0008047 (2000).

\bibitem{RESI1}
M.~Reed and B.~Simon.
\newblock \emph{Methods of modern mathematical physics. {I}}.
\newblock Academic Press, San Diego (1980).

\bibitem{vNeuP}
J.~von Neumann.
\newblock \emph{On infinite direct products}.
\newblock Compos. Math. \textbf{6}, 1--77 (1938).
\newblock cf. also Collected Works {III}, No. 6.

\bibitem{BraRob}
O.~Bratteli and D.~W. Robinson.
\newblock \emph{Operator algebras and quantum statistical mechanics. {I+II}}.
\newblock Springer, New York (1979, 1997).

\bibitem{Oza}
M.~Ozawa.
\newblock \emph{Measuring processes and repeatability hypothesis}.
\newblock In \emph{Probability theory and mathematical statistics (Kyoto,
  1986)}, volume 1299 of \emph{Lect. Notes Math.}, pages 412--421. Springer,
  Berlin (1988).

\bibitem{QHA2}
R.F. Werner.
\newblock \emph{Physical uniformities on the state space of non-relativistic
  quantum mechanics}.
\newblock Found. Phys. \textbf{13}, 859--881 (1983).

\bibitem{Tak}
M.~Takesaki.
\newblock \emph{Theory of operator algebras}.
\newblock Springer, New York, Heidelberg, Berlin (1979).

\bibitem{AraWoo}
H.~Araki and E.J. Woods.
\newblock \emph{A classification of factors}.
\newblock Publ. R.I.M.S, Kyoto Univ. \textbf{4}, 51--130 (1968).

\bibitem{connes}
A.~Connes.
\newblock \emph{Sur la cassification des facteurs de type {II}}.
\newblock C.R. Acad. Sci. Paris Ser. \textbf{A-B 281}, A13--A15 (1975).

\bibitem{Werner89}
R.~F. Werner.
\newblock \emph{Quantum states with {Einstein-Podolsky-Rosen} correlations
  admitting a hidden-variable model}.
\newblock Phys. Rev. \textbf{A 40}, no.~8, 4277--4281 (1989).

\bibitem{Cirelson}
B.S. Cirel'son.
\newblock \emph{Quantum generalizations of {Bell's} inequalities}.
\newblock Lett. Math. Phys. \textbf{4}, 93--100 (1980).

\bibitem{SumWer95}
S.J. Summers and R.F.Werner.
\newblock \emph{On {Bell's} inequalities and algebraic invariants}.
\newblock Lett. Math. Phys. \textbf{33}, 321--334 (1995).

\bibitem{BellRev}
R.~F. Werner and M.~M. Wolf.
\newblock \emph{Bell inequalities and entanglement}.
\newblock Quant. Inf. Comp. \textbf{1}, no.~3, 1--25 (2001).

\bibitem{Tak70}
M.~Takesaki.
\newblock \emph{Tomita's theory of modular Hilbert algebras and its
  application}, volume 128 of \emph{Lect. Notes. Math.}
\newblock Springer, Berlin, Heidelberg, New York (1970).

\bibitem{KMS}
R.~Haag, N.M. Hugenholtz and M.~Winnink.
\newblock \emph{On the equilibrium states in quantum statistical mechanics}.
\newblock Commun. Math. Phys. \textbf{5}, 215--236 (1967).

\bibitem{AV}
R.~Arens and V.S. Varadarajan.
\newblock \emph{On the concept of {EPR} states and their structure}.
\newblock Jour. Math. Phys. \textbf{41}, 638--651 (2000).

\bibitem{WAV}
R.F. Werner.
\newblock \emph{{EPR} states for von {Neumann} algebras}.
\newblock quant-ph/9910077 (1999).

\bibitem{Sak}
S.~Sakai.
\newblock \emph{{C*}-algebras and {W*}-algebras}.
\newblock Springer, Berlin, Heidelberg, New York (1971).

\bibitem{BaumWoll}
H.~Baumg{\"a}rtel and M.~Wollenberg.
\newblock \emph{Causal nets of operator algebras}.
\newblock Akademie Verlag, Berlin (1992).

\bibitem{BraEll92}
D.E.~Eveans O.~Bratelli, G.A.~Elliott and A.~Kishimoto.
\newblock \emph{Non-commutative spheres {II}: Rational rotations}.
\newblock J. Operator Theory \textbf{27}, 53--85 (1992).

\bibitem{HoegSkjel81}
R.~H{\o}egh-Krohn and T.~Skjelbred.
\newblock \emph{Classification of {C*}-algebras admitting ergodic actions of
  the two-dimensional torus}.
\newblock J. Reine Angew. Math. \textbf{328}, 1--8 (1981).

\bibitem{KorLeu}
N.~Korolkova and G.~Leuchs.
\newblock \emph{Multimode quantum correlations}.
\newblock In \emph{Coherence and statistics of photons and atoms} ( J.~Perina,
  editor). Wiley (2001).

\bibitem{Zeil02}
C.~Brukner, M.S. Kim, J-W. Pan and A.~Zeilinger.
\newblock \emph{Correspondence between continuous variable and discrete quantum
  systems of arbitrary dimensions}.
\newblock quant-ph/0208116 (2002).

\bibitem{NeuDipl}
M.~Neumann.
\newblock \emph{Verletzung der {Bellschen} {Ungleichungen} f{\"u}r
  {Gau{\ss}sche} {Zust{\"a}nde}}.
\newblock Diplomarbeit, TU-Braunschweig (2002).

\bibitem{Chandler}
R.E. Chandler.
\newblock \emph{Hausdorff compactifications}, volume~23 of \emph{Lect. Notes
  Pure Appl. Math.}
\newblock Dekker, New York (1976).

\bibitem{Alfsen}
E.M. Alfsen.
\newblock \emph{Compact convex sets and boundary integrals}, volume~57 of
  \emph{Ergebnisse der Mathematik und ihrer Grenzgebiete}.
\newblock Springer, Springer, New York, Hedelberg, Berlin, (1971).

\end{thebibliography}

\end{document}